\documentclass[12pt]{article}
\usepackage{graphicx,slashed,multicol}
\makeatletter
\newcounter{alphaequation}[equation]
\def\thealphaequation{\theequation\hbox to
0.6em{\hfil\alph{alphaequation}\hfil}}
\def\eqnsystem#1{
\def\@eqnnum{{\rm (\thealphaequation)}}
\def\@@eqncr{\let\@tempa\relax \ifcase\@eqcnt \def\@tempa{& & &} \or
\def\@tempa{& &}\or \def\@tempa{&}\fi\@tempa
\if@eqnsw\@eqnnum\refstepcounter{alphaequation}\fi
\global\@eqnswtrue\global\@eqcnt=0\cr}
\refstepcounter{equation} \let\@currentlabel\theequation \def\@tempb{#1}
\ifx\@tempb\empty\else\label{#1}\fi
\refstepcounter{alphaequation}
\let\@currentlabel\thealphaequation
\global\@eqnswtrue\global\@eqcnt=0 \tabskip\@centering\let\\=\@eqncr
$$\halign to \displaywidth\bgroup \@eqnsel\hskip\@centering
$\displaystyle\tabskip\z@{##}$&\global\@eqcnt\@ne
\hskip2\arraycolsep\hfil${##}$\hfil& \global\@eqcnt\tw@\hskip2\arraycolsep
$\displaystyle\tabskip\z@{##}$\hfil
\tabskip\@centering&\llap{##}\tabskip\z@\cr}
\def\endeqnsystem{\@@eqncr\egroup$$\global\@ignoretrue} \makeatother

\font\tenrsfs=rsfs10 at 12pt
\font\sevenrsfs=rsfs7
\font\fiversfs=rsfs5
\newfam\rsfsfam
\textfont\rsfsfam=\tenrsfs
\scriptfont\rsfsfam=\sevenrsfs
\scriptscriptfont\rsfsfam=\fiversfs
\def\mathscr#1{{\fam\rsfsfam\relax#1}}
\def\Lag{\mathscr{L}}
\makeatletter
\def\art{\@ifnextchar[{\eart}{\oart}}
\def\eart[#1]#2#3#4#5#6{{\rm #2}, {\em #3 \rm#4} {\rm (#6) #5} ({\em arXiv:#1})}
\def\hepart[#1]#2{{\rm #2, \em arXiv:#1}}
\newcommand{\oart}[5]{{\rm #1}, {\em #2 \bf #3} {\rm (#5) #4}}

\newcommand{\be}{\begin{equation}}
\newcommand{\ee}{\end{equation}}

\def\bsg{\ifmmode B\to X_s\gamma\else $B\to X_s\gamma$\fi}
\def\bsll{\ifmmode B\to X_s\ell^+\ell^-\else $B\to X_s\ell^+\ell^-$\fi}
\def\shat{\ifmmode \hat{s}\else $\hat{s}$\fi}

\newcommand{\newc}{\newcommand}

\newc{\gsim}{\lower.7ex\hbox{$\;\stackrel{\textstyle>}{\sim}\;$}}
\newc{\lsim}{\lower.7ex\hbox{$\;\stackrel{\textstyle<}{\sim}\;$}}
\newc{\ie}{{\it i.e.}}
\newc{\etal}{{\it et al.}}
\newc{\mev}{\hbox{\rm\,MeV}}
\newc{\gev}{\hbox{\rm\,GeV}}
\newc{\tev}{\hbox{\rm\,TeV}}
\newc{\xpb}{\hbox{\rm\, pb}}
\newc{\xfb}{\hbox{\rm\, fb}}

\newc{\mtop}{m_t}
\newc{\mbot}{m_b}
\newc{\mz}{M_Z}
\newc{\mw}{M_W}
\newc{\alphasmz}{\alpha_s(M_Z)}
\newc{\swsq}{\sin^2\theta_W}
\newc{\cwsq}{\cos^2\theta_W}
\newc{\tw}{\tan\theta_W}
\newc{\cw}{\cos\theta_W}
\newc{\sw}{\sin\theta_W}
\newc{\BR}{\hbox{\rm BR}}
\newc{\zbb}{Z\to b\bar}
\newc{\Gb}{\Gamma (Z\to b\bar b)}
\newc{\Gh}{\Gamma (Z\to \hbox{\rm hadrons})}
\newc{\sgn}{\mbox{sgn}}

\def\eq#1{eq.~(\ref{#1})}
\def\fig#1{fig.~\ref{#1}}
\def\slash#1{\not  \! {#1}}

\def\vev#1{\langle {#1} \rangle}

\newcounter{mysubequation}[equation]

\newcommand{\TeV}{\,\mathrm{TeV}}
\newcommand{\GeV}{\,\mathrm{GeV}}

\def\beq{\begin{equation}}
\def\eeq{\end{equation}}
\def\bea{\begin{eqnarray}}
\def\eea{\end{eqnarray}}
\def\slashchar#1{\setbox0=\hbox{$#1$}           
   \dimen0=\wd0                                 
   \setbox1=\hbox{/} \dimen1=\wd1               
   \ifdim\dimen0>\dimen1                        
      \rlap{\hbox to \dimen0{\hfil/\hfil}}      
      #1                                        
   \else                                        
      \rlap{\hbox to \dimen1{\hfil$#1$\hfil}}   
      /                                         
   \fi}                                         %
\catcode`@=11
\long\def\@caption#1[#2]#3{\par\addcontentsline{\csname
  ext@#1\endcsname}{#1}{\protect\numberline{\csname
  the#1\endcsname}{\ignorespaces #2}}\begingroup
    \small
    \@parboxrestore
    \@makecaption{\csname fnum@#1\endcsname}{\ignorespaces #3}\par
  \endgroup}
\catcode`@=12

\begin{document}

\baselineskip=15pt

\setcounter{footnote}{0}
\setcounter{figure}{0}
\setcounter{table}{0}

\begin{titlepage}
\begin{flushright}
CERN-PH-TH/2010-049\\
IFUP-TH/2010-1
\end{flushright}
\vspace{.3in}
\begin{center}
{\Large \bf Lorentz Violation from the Higgs Portal}

\vspace{0.5cm}

{\bf Gian F. Giudice$^a$, Martti Raidal$^{b,c}$, Alessandro Strumia$^{a,d}$}

\vspace{.5cm}

{\it (a) CERN, Theory Division, CERN, CH-1211 Geneva 23, Switzerland}\\
{\it (b) National Institute of Chemical Physics and Biophysics, Ravala 10, Tallin, Estonia}\\
{\it (c) Department of Physics, P.O.Box 64, FIN-00014 University of Helsinki}\\
{\it (d) Dipartimento di Fisica dell'Universit\`a di Pisa and INFN, Italia}

\end{center}
\vspace{.8cm}

\begin{abstract}
\medskip
We study bounds and signatures of models where the Higgs doublet has an inhomogeneous mass or vacuum expectation value, 
being coupled to a hidden sector that breaks Lorentz invariance.
This physics is best described by a low-energy effective Lagrangian in which the Higgs speed-of-light is smaller than $c$;
such effect is naturally small because it is suppressed by four powers of the inhomogeneity scale. The Lorentz violation in the Higgs sector is communicated at tree level to fermions (via Yukawa interactions) and to massive gauge bosons, although the most important effect comes from one-loop diagrams for photons and from two-loop diagrams for fermions.
We calculate these effects by deriving the renormalization-group equations for the speed-of-light of the Standard Model particles.
An interesting feature is that the strong coupling dynamically makes the speed-of-light equal for all colored particles.

\end{abstract}

\bigskip
\bigskip


\end{titlepage}


\section{Introduction}
\label{intro}

The sector responsible for the electroweak symmetry breaking still leaves open theoretical questions and is experimentally unknown. Its most plausible explanation relies on the idea of spontaneously broken gauge symmetry, although the Higgs mechanism introduces its own problems. The main puzzle is associated with the presence of a mass parameter for the Higgs field, which sets the scale for the electroweak phenomena. At the quantum level this mass term is quadratically sensitive to short-distance physics. Actually, being the only super-renormalizable interaction in the Standard Model, this mass term can be viewed as a window open towards the influence of new and unknown high-energy  or hidden sectors of the theory.  New scalars $M(x)$, neutral under the SM gauge group, can have renormalizable couplings to the Higgs $H$:
\beq M^2(x) |H|^2 . \label{M(x)}\eeq
This aspect was discussed by several authors and was dubbed ``Higgs portal" in~\cite{wil}. 
In this paper we consider the possibility that this Higgs portal connects the Standard Model with some hypothetical sector that breaks Lorentz invariance, 
such that $M(x)$ has a space-time dependent vacuum expectation value (vev) varying on a characteristic small length-scale $\ell$.

Violation of Lorentz invariance is not uncommon in certain theories of quantum gravity, as in the presence of a space-time foam, and even in string theory. 
Alternatively the Higgs field itself might be `foamy', existing only in tiny islands of space-time.
Or maybe its vev might be `foamy', being non-zero only in some regions,
giving rise to a small average vev from a larger fundamental vev.
In both cases an apparently constant Higgs vev is obtained at low energy, i.e.\ after averaging over length-scales much bigger than $\ell$.
Here we study the low energy signals of these kinds of scenarios, performing concrete computations from the interaction
in \eq{M(x)}.

In section 2 we compute the Lorentz non-invarant
dispersion relation satisfied by a scalar or a fermion with a non-constant mass $M(x)$.
In section 3 we develop a general technique to obtain the full effective Lagrangian.
In section 4 we write RGE equations for the speed-of-light of the various SM particles, finding 
how Lorentz-breaking in the Higgs sector propagates at loop level to all other particles, and how
a strong coupling can dynamically restore the Lorentz symmetry.
In section 5 we consider the signals and constraints, and in section 6 we conclude.

\section{Propagation of particles with space dependent masses}
\label{secprop}
One of the consequences of the scenario we consider is a non-constant Higgs vev, and consequently a non-constant mass for SM particles.
In order to obtain some physical intuition about our setting, we start by considering the propagation of a complex scalar particle or of a Dirac fermion with masses that vary periodically in space. 

\subsection{Scalar}
We study the case of a complex scalar $H(x)$ with a squared mass $M^2(x)$ that depends only on one spatial coordinate $x$.
We assume that $M^2(x)$ has period $\ell$, and  constant values $M_1^2$ and $M_2^2$
within intervals of length $r_1\ell$ and $r_2\ell$ (with $0<r_{1,2}<1$ and $r_1+r_2=1$):
\beq
M^2(x)= \left\{ \begin{array}{ll}
M_1^2 ~~~&{\rm for} ~0<x~({\rm mod}~\ell)<r_1\ell \\ M_2^2~~~&{\rm for} ~r_1\ell< x~({\rm mod}~\ell)<\ell
\end{array} \right. .
\eeq

According to the Floquet-Bloch theorem~\cite{Floquet}, as a result of the periodicity of $M^2(x)$, the solution of the Klein-Gordon equation is of the form
\beq
H(x,t) = e^{-i(Et-kx)}u(x)~,
\eeq
where $u(x)$ also has periodicity $\ell$. The Klein-Gordon equation for $u(x)$ is given by
\beq
\left[ \frac{d^2}{dx^2}+2ik\frac{d}{dx} +E^2-k^2-M^2(x)\right] u(x)=0~ ,
\eeq
and has the solution
\beq
u(x)= \left\{ \begin{array}{ll}
A_1 e^{i(k_1-k)x}+B_1 e^{-i(k_1+k)x} ~~~&{\rm for} ~0<x<r_1\ell \\
A_2 e^{i(k_2-k)x}+B_2 e^{-i(k_2+k)x} ~~~&{\rm for} ~r_1\ell<x<\ell
\end{array} \right. ,
\eeq
\beq
k_{1,2}\equiv \sqrt{E^2-M_{1,2}^2}~.
\label{defk}
\eeq
Continuity of the function $u(x)$ and of its first derivative at the matching points $x=0$ and $x=r_1\ell$ imposes four constraints. Three of them determine the integration constants $A_{1,2}$ and $B_{1,2}$ up to an overall normalization, while the fourth equation defines the dispersion relation:
\beq
\cos (k\ell)=\cos (k_1r_1\ell)\cos(k_2r_2\ell)-\frac{k_1^2+k_2^2}{2k_1k_2}\sin(k_1r_1\ell)\sin(k_2r_2\ell).
\label{dispsc}
\eeq
This equation describes the relation between energy $E$ and momentum $k$.
We see that $k$ is fixed up to a $2\pi /\ell$ ambiguity.

Since we are assuming that Lorentz violation is related to phenomena at very short distance, we are interested in particle propagation for momenta much smaller than $1/\ell$. In this limit, the dispersion relation in \eq{dispsc} can be written in the familiar form
\beq
E^2=k^2c^2+m^2c^4,
\label{dispfam}
\eeq
where
\beq
m^2\equiv M_1^2r_1+M_2^2r_2 +{\cal O}(\ell^2),~~~c\equiv 1-\frac{r_1^2r_2^2 ( 1+2r_1r_2)}{360} (M_1^2-M_2^2)^2\ell^4+{\cal O}(\ell^6).
\label{result}
\eeq
Therefore, when the particle is observed at momenta much smaller than $1/\ell$, the effect of a space-varying mass can be absorbed in a redefinition of its mass and of its ``light speed" (or, more appropriately, of the maximal attainable velocity in the massless limit). While the mass redefinition is unobservable (unless we have a theory in which particle masses can be predicted), the redefinition of the ``light speed" can be experimentally measured when the particle propagation is compared with another particle with different value of $c$. Thus, a scalar particle with a space-varying mass, when viewed at low energies, appears as a particle with a  constant mass, given by the square root of the average of $M^2(x)$, 
but with a modified relation between energy and momentum. 

The Lorentz-violating effect trivially disappears when $M_1^2\to M_2^2 $, since the source of Lorentz violation vanishes in this limit. More interestingly, Lorentz violation
also disappears when $\ell\to 0$. In this limit, the characteristic length of the mass variation becomes infinitely smaller than the de Broglie wavelength of the particle. Since the source of Lorentz  violation $M^2(x)$ has dimension of mass squared,
the adimensional correction to $c$ must be suppressed by the high scale $\Lambda = 2\pi/\ell$.
Thereby, in our scenario, high-scale physics 
generates small Lorentz-breaking effects.
This is unlike a generic Lorentz-breaking scenario (such as `space-time foam'), where one typically expects order unity deviations from $c=1$
even from Lorentz-violation at the Planck scale.\footnote{Phenomenological
analyses assume that $c=1$ and focus on effects from higher-dimensional operators that grow with some unknown power of energy,
although such effects at loop level also give rise to a power-divergent correction to $c$,
and more generically to some of the Lorentz-violating operators of~\cite{Kost}.
In our case we instead neglect effects that grow with energy, because they are suppressed by more powers of $\ell$.
}
The correction to $c$ in \eq{result} is always negative and thus the ``light speed" of a scalar is smaller than the canonical value.

\subsection{Fermion}
We can now repeat the discussion in the case of a fermion. Suppose that its Dirac mass
depends on $x$ with period $\ell$, being constant within intervals of lengths $r_1\ell$ and $r_2\ell$
\beq
M(x)= \left\{ \begin{array}{ll}
M_1 ~~~&{\rm for} ~0<x~({\rm mod}~\ell)<r_1\ell\\ M_2~~~&{\rm for} ~r_1\ell<x~({\rm mod}~\ell)<\ell
\end{array} \right. .
\eeq
Again, we are considering only one space dimension. Exploiting the Floquet-Bloch theorem, we can decompose the fermion field as
\beq
\psi (x,t)=e^{-i(Et-kx)}\left( \begin{array}{llll}
u_1^{(+)}(x) \\ u_2^{(-)}(x) \\ u_1^{(-)}(x) \\ u_2^{(+)}(x)
\end{array} \right) ,
\eeq
where the 2-component spinors $u^{(\pm )}$ are periodic functions. In the Weyl basis, the Dirac equation becomes
\bea
\left( i \frac{d}{dx} +E-k\right) u^{(+)}(x) &=&-M(x) u^{(-)}(x) \nonumber \\
\left( i \frac{d}{dx} -E-k\right) u^{(-)}(x) &=&M(x) u^{(+)}(x) .
\eea 
The solution is
\begin{eqnsystem}{sys:Dir}
u^{(+)}(x)&=& \left\{ \begin{array}{ll}
A_1 e^{i(k_1-k)x}-\frac{B_1 M_1}{k_1+E} e^{-i(k_1+k)x}\phantom{-} ~~~&{\rm for} ~0<x<r_1\ell\\
A_2 e^{i(k_2-k)x}- \frac{B_2 M_2}{k_2+E}e^{-i(k_2+k)x} ~~~&{\rm for} ~r_1\ell<x<\ell
\end{array} \right. \\
u^{(-)}(x)&=& \left\{ \begin{array}{ll}
-\frac{A_1 M_1}{k_1+E} e^{i(k_1-k)x}+B_1  e^{-i(k_1+k)x} ~~~&{\rm for} ~0<x<r_1\ell\\
-\frac{A_2 M_2}{k_2+E} e^{i(k_2-k)x}+B_2  e^{-i(k_2+k)x} ~~~&{\rm for} ~r_1\ell<x<\ell
\end{array} \right. ,
\end{eqnsystem}
where $k_{1,2}$ are again given by \eq{defk}.
The conditions of continuity of the functions $u^{(\pm)}$ at the matching points $x=0$ and $x=r_1\ell$ determine, up to their normalization, the integration constants $A_{1,2}$ and $B_{1,2}$ and the dispersion relation
\beq
\cos (k\ell)=\cos (k_1r_1\ell)\cos(k_2r_2\ell)+\frac{M_1M_2-E^2}{k_1k_2}\sin(k_1r_1\ell)\sin(k_2r_2\ell).
\label{dispfr}
\eeq
In the limit of small momenta, the 
energy-momentum relation is given by the familiar expression $E^2=k^2c^2+m^2c^4$, where
\beq
m\equiv M_1r_1+M_2r_2 +{\cal O}(\ell^2),~~~c\equiv 1-\frac{r_1^2r_2^2}{6}(M_1-M_2)^2\ell^2+{\cal O}(\ell^4).
\label{bubu}
\eeq
Again, at small momenta, the total effect of the space-varying mass can be parametrized by a distortion of the ``light speed". Notice that the ``light speed" of the scalar deviates from the canonical value at order $\ell^4$ while, for space-varying fermion masses, the effect comes already at order $\ell^2$. This is because the Lorentz-violating effect is always suppressed by two powers of the mass inhomogeneity, which amounts to $(M_1^2-M_2^2)^2$ for the scalar and $(M_1-M_2)^2$ for the fermion. Dimensional arguments then determine the different powers of $\ell$. As in the case of the scalar, the correction to $c$ in \eq{bubu} is negative and thus the maximal attainable velocity is smaller than the ordinary light speed.

\section{Low-energy effective theory}
\label{seceff}

After having clarified the physical meaning of a particle with space-varying mass, we can proceed in the analysis of the Standard Model with a Lorentz-violating Higgs mass parameter. Our goal is to construct an effective theory valid at energies below a cutoff scale $\Lambda$, obtained by integrating out the high-frequency modes. Here $1/\Lambda$ represents the typical length of the variations of the Higgs mass. Since $\Lambda$ is the energy scale at which the Lorentz violation, originating in a hidden sector, is communicated to the Higgs field, we assume that $\Lambda$ is much larger than the TeV scale.  The space dependent mass $M(x)$ mixes the low-frequency modes (with Fourier momentum $k\ll \Lambda$)
with the high-frequency ones.  By integrating out the high-frequency modes, their effects is described at low energy by
Lorentz-violating operators.
Let us explain the procedure to derive the effective theory.

\smallskip

\subsection{Higgs effective Lagrangian}

We denote with $H$ the Higgs doublet and introduce in the Lagrangian a space-time dependent component for its mass,
\beq
\Lag=-H^\dagger (x) \left[ \partial^2 +M^2+\mu^2  F(\hat x )\right]H(x).
\label{higgsl}
\eeq
Here $M$ is a mass parameter of the order of the electroweak scale, $\mu$ is a mass parameter (much smaller than the electroweak scale) parametrizing the amplitude of the space-time varying component, and $F$ is a
generic order unity dimensionless function that modulates the space-time dependence. For simplicity we take $F$ to depend on a single combination of space-time coordinates,
\beq
 \hat x \equiv \frac{x\cdot a}{-a^2},
 \eeq
 where $a$ is a fixed 4-vector. It is convenient to normalize $\hat x$ with $-a^2$ because we have in mind a space-like fluctuation of the Higgs mass ($a^2<0$), but our results remain valid also for time-like variations ($a^2>0$).
 We assume that $F$ is a real function with the following three properties. {\it (i)} It is periodic: $F(\hat x +2\pi n)=F(\hat x)$ for any integer $n$.  {\it (ii)} It is bounded: $|F( \hat x )|\le 1$. {\it (iii)} It averages to zero within one period: $\int_0^{2\pi} d\hat x F(\hat x ) =0$.
 
Being periodic, the function $F$ can be expanded in an infinite sum of Fourier modes,
\beq
F(\hat x )=\sum_{n=-\infty}^{+\infty} f_n e^{in \hat x }~~~~{\rm with}~ f_n\equiv \frac{1}{2\pi}\int_0^{2\pi}d\hat x \, F(\hat x )e^{-in \hat x }.
\eeq
The Fourier coefficients $f_n$ are such that $f_n^* =f_{-n}$, because $F$ is real, and such that $f_0=0$, because of the property {\it (iii)} above. Moreover, the $f_n$ are real (imaginary) when $F$ is an even (odd) function of $\hat x$.

It is convenient to work in Fourier space and express the Higgs field $H (x)$ as
\beq
H (x) =\frac{1}{(2\pi)^2} \int d^4k\, e^{ikx} H(k).
\eeq
We can decompose the quadri-momentum $k$ in terms of the quantities
\beq
k_\parallel \equiv \frac{k\cdot a}{|a|},\qquad
k_\perp \equiv k-\frac{k \cdot a}{a^2} a,\qquad
|a| \equiv \sqrt{-a^2},
\eeq
which have been defined such that $k_\perp \cdot a =0$ and $k^2=k_\perp^2-k_\parallel^2$. Note that $k_\parallel^2$ is positive (negative) for space-like (time-like) fluctuations of the Higgs mass.
The integration over $k_\parallel $ of a generic function $g(k)$ can be decomposed into an infinite sum of integrations within shells of momenta $(n-1/2)/|a|<k_\parallel <(n+1/2)/|a|$ for any arbitrary integer $n$:
\beq
\int_{-\infty}^{+\infty}dk_\parallel \, g(k_\perp,k_\parallel  )=\sum_{n=-\infty}^{+\infty} \int_{-(2|a|)^{-1}}^{(2|a|)^{-1}}dk_\parallel  \, g\left( k_\perp,k_\parallel  +\frac{n}{|a|}\right) .
\eeq
Using this expansion, the Higgs action in \eq{higgsl} becomes
$$
\mathscr{S}= \int d^3k_\perp \int_{-(2|a|)^{-1}}^{(2|a|)^{-1}} dk_\parallel   \left\{  \sum_{n=-\infty}^{+\infty}
H^\dagger(k_\parallel  +\frac{n}{|a|})\left[ k_\perp^2 - \left(k_\parallel  +\frac{n}{|a|}\right)^2-M^2\right] H(k_\parallel  +\frac{n}{|a|}) \right. \nonumber 
$$
\beq
\left. -\mu^2\sum_{n,m=-\infty}^{+\infty}f_{n-m}H^\dagger(k_\parallel  +\frac{n}{|a|})H(k_\parallel  +\frac{m}{|a|})\right\},
\label{higgslf}
\eeq
where the dependence of $H$ on $k_\perp$ is understood.

The first line of \eq{higgslf} is just the usual SM Lagrangian, that gives the ordinary Lorentz-invariant dispersion relation for each mode with momentum $k$.
The term proportional to $\mu^2$ in the second line of \eq{higgslf} introduces a mixing between the different modes of the Higgs field. In particular, the zero mode of the Higgs field ($n=0$) mixes with every high-frequency mode $n$ with a coefficient $\mu^2 f_n$. Notice that the term proportional to $\mu^2$ generates only off-diagonal mixings, since $f_0=0$.
A non-periodic $M^2(x)$ would give a continuous (rather than discrete) mixing, but the final result would be the same as long
as the Fourier transform of $M^2(x)$ vanishes fast enough at $k\to 0$, so that 
an unambiguous splitting between low-momentum and high-momentum modes still exists.

Coming back to the periodic $M^2(x)$,
the low-energy effective theory is obtained by integrating out all modes of the Higgs field $H$ with $n\ne 0$. This procedure leads to a non-trivial result because of the mixing of the high-frequency modes with the zero mode. To obtain the effective theory, it is convenient to express the high-frequency modes through their equations of motion at first order in $\mu^2$,
\beq
H (k_\parallel  +\frac{n}{|a|})=\frac{\mu^2 f_n}{k_\perp^2-\left( k_\parallel  +\frac{n}{|a|}\right)^2-M^2}H(k_\parallel ) + \mbox {high-frequency terms}.
\label{eqmo}
\eeq
Replacing \eq{eqmo} into \eq{higgslf}, retaining only terms involving zero modes ($n=0$) and expanding the result for small $|a|$, we  obtain the low-energy effective theory for the Higgs field: 
\beq
\mathscr{S} _{\rm eff}=  \int d^4k \, H^\dagger (k) \left[ Z(k^2-M^2) -\Delta M^2 -2\, \delta c~  {k_\parallel }^2 \right] H(k),
\label{effecth}
\eeq
\beq
Z= 1-\frac{\delta c}{2},~~~ \Delta M^2= 2\mu^4a^2\sum_{n=1}^{\infty}\frac{|f_n|^2}{n^2},~~~\delta c = -4\mu^4a^4\sum_{n=1}^{\infty}\frac{|f_n|^2}{n^4}.
\label{epsge}
\eeq
In \eq{effecth}, the integration is only over momenta $k$ smaller than the cutoff  $\Lambda =1/|a|$.

While $Z$ and $\Delta M^2$ can be absorbed in the wave-function and mass definitions, $\delta c$ leads to a physical effect. Note that, for space-like variations of the Higgs mass ($a^2<0$), $\Delta M^2$ is negative and one could imagine scenarios in which the electroweak breaking is triggered solely by Lorentz-violating effects.
 
In coordinate space, the new effect is described by a  Lorentz-violating term:
\beq
\mathscr{S}_{\rm eff} =  \int d^4x\ \Lag_{\rm eff},\qquad
\Lag_{\rm eff} = |\partial_\mu H|^2 - M^2 |H|^2- 2\, \delta c  
\, H^\dagger \frac{( a\cdot  \partial )^2}{a^2} H .
\label{effo}
\eeq
This is a renormalizable interaction. Its coefficient 
$\delta c \sim \mu^4/\Lambda^4$ is however suppressed by four powers of the cutoff scale, at which the Lorentz violation is communicated to the Higgs sector. This operator modifies
the Higgs kinetic term in such a way that the ``light speed" for the Higgs field along the direction identified by the quadri-vector $ a$ becomes
\beq
c = 1+\delta c.
\eeq

By making specific assumptions on the function $F(\hat x)$ that modulates the space-time dependence of the Higgs mass, we can explicitly calculate the expression of $\Delta M^2$ and $\delta c$ from \eq{epsge}. For instance, if $F(\hat x)= \cos (\hat x )$, the Fourier coefficients are $f_{\pm 1}=1/2$ and $f_n =0$ for $n\ne \pm 1$. Hence, we obtain
\beq
\Delta M^2 =-\frac{\mu^4}{2\Lambda^2},\qquad \delta c =- \frac{\mu^4}{\Lambda^4} ~~~{\rm for}~F(\hat x)= \cos (\hat x ).
\eeq
Another example is the square-wave function
\beq
F(\hat x)=\left\{ \begin{array}{ll}
+1~~~&{\rm for}~2n\pi <\hat x< (2n+1)\pi \\
-1~~~&{\rm for}~(2n+1)\pi <\hat x< (2n+2)\pi
\end{array} \right. .
\eeq
In this case $f_n=i[(-1)^n-1]/(\pi n)$ and thus we obtain
\beq
\Delta M^2 =-\frac{\pi^2\mu^4}{12\Lambda^2},\qquad \delta c = -\frac{\pi^4\mu^4}{60\Lambda^4} ~~~{\rm for~square~wave}~F(\hat x) .
\eeq
Notice that this expression of $\delta c$ coincides with the result in \eq{result} obtained by solving the Klein-Gordon equation with variable mass,  after the replacement $r_{1,2}=1/2$, $\ell= 2\pi/\Lambda$ and $|M_1^2-M_2^2| = 2\mu^2$. 

So far we have studied the case in which the Lorentz violation identifies one special direction in space-time, but our results can be easily generalized. Actually the derivation of the effective theory used a generic 4-vector $a$ and can be adapted to different cases. For instance, taking a 4-vector $a$ with vanishing space components corresponds to the 
rotationally invariant case, which leads to a Lorentz-violating Lagrangian term
\beq
2\, \delta c  \,  H^\dagger {\vec \nabla}^2 H ,\qquad \delta c \sim -\mu^4/\Lambda^4.
\label{effo2}
\eeq 
When $M^2(x)$ has the symmetry of a cube (corresponding to the octahedral group), the dimension-4 effective operator is still of the form of \eq{effo2}, exhibiting rotational symmetry. This is
because the usual $\delta_{ij}$ is the only two-index invariant tensor of both the octahedral group and the full SO(3) rotation group. The breaking of the rotational symmetry will appear only in
higher dimensional operators. Another case leading to \eq{effo2} is the one in which $M^2(x)$ is a randomly-varying function. This gives the rotationally invariant effective operator, just like the random motion of molecules gives, on average,
a rotationally invariant refraction index of air.
In the following, just for simplicity, we focus on the rotationally-symmetric case, which contains all the important features of Higgs-induced Lorentz violation. Finally note that, after taking into account gauge corrections, the operator in \eq{effo2} gets gauge-covariantized
as usual, $\vec \nabla \to \vec D = \vec \nabla + i g \vec A$.

\subsection{Full effective Lagrangian}

\begin{figure}[t]
$$\includegraphics{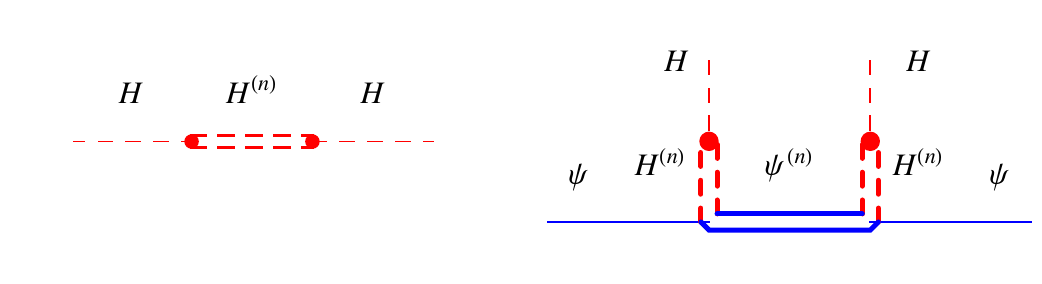}$$
\caption{\em The diagrams generating the effective Lorentz-violating interactions, obtained after integrating out the high-frequency modes, for the terms with two Higgs bosons ({\it a}) and two fermions ({\it b}). Single lines denote propagators of low-frequency modes for the Higgs boson (dashed line) and the fermion (solid line). Double lines denote propagators of the high-frequency modes, as given by \eq{fe33} for the Higgs boson and by \eq{fe35} for the fermion. The dot denotes the mixing between high- and low-frequency modes, as given by \eq{fe34}.
\label{Feyn1}}
\end{figure}

The construction of the low-energy effective Lagrangian for the Higgs field can now be extended to the full Standard Model. Each Standard Model field is expanded in Fourier space and the high-frequency modes are integrated out. This procedure can be carried out with the help of the equations of motion, as discussed above, or, more simply, with the Feynman diagram technique. The propagator of the $n$-mode Higgs field, expanded for small $|a|$, is given by
\beq
\frac{ia^2}{n^2}\left[ 1-2k_\parallel  \frac{|a|}{n} -(4{k_\parallel }^2+ k^2-M^2)\frac{a^2}{n^2} + {\cal O}(a^3) \right]~.
\label{fe33}
\eeq
The mixing between the zero-mode and any $n$-mode of the Higgs field corresponds to a mass insertion
\beq
i \mu^2 f_n ~.
\label{fe34}
\eeq
A summation $\sum_{n=-\infty}^{+\infty}$ is required in diagrams with high-frequency modes in the internal lines.

Using these rules, we can easily recover the Lagrangian in \eq{effecth} from the Feynman diagram in fig.~1a. We can then extend the calculation to the other Standard Model fields. The Lorentz violation, originally residing in the Higgs mass term, is communicated to quarks and leptons through the tree-level diagram of fig.~1b. The propagator of the $n$-mode fermion field, expanded for small $|a|$, is given by
\beq
\frac{i \slash a}{n} + \frac{ia^2}{n^2} (\slash k +m) - \frac{2i |a| \slash a k_\parallel}{n^2}  + {\cal O}(a^3) ~,
\label{fe35}
\eeq
where $m$ is the mass of the fermion field.
Hence, we obtain that the diagram in fig.~1b generates the following effective operator connecting two Higgs ($H$) and two fermion fields ($\psi$),
\beq
i \epsilon_\psi~ (H^\dagger \bar\psi) \frac{\slash a a\cdot \partial}{a^2} (\psi H) ~,
\label{opfer}
\qquad
\epsilon_\psi =12 \lambda^\dagger \lambda \mu^4 a^6 \sum_{n=1}^{\infty} \frac{|f_n|^2}{n^6} ~.
\eeq
Here $\lambda$ is the corresponding Yukawa coupling. 
If $\psi$ is a weak singlet, contractions of  SU$(2)_L$ indices give $|H|^2$;
if $\psi$ is a weak doublet, $ \psi \vev{H}$ is the component of $\psi$ that gets mass from the Yukawa $\lambda$.

In the special cases in which the modulating function $F(\hat x)$ is a cosine or a square wave, $\epsilon_\psi$ becomes
\beq
\epsilon_\psi =  -\lambda^\dagger \lambda \frac{\mu^4}{\Lambda^6} \times
\left\{\begin{array}{ll}
3 ~~~&{\rm for}~F(\hat x)= \cos (\hat x )\cr
  {51\pi^6}/{10080} &{\rm for~square~wave}~F(\hat x) 
  \end{array}\right. \ .
\eeq

The operator in \eq{opfer}, after electroweak symmetry breaking, gives a modification of the ``light speed" of the fermion along the direction identified by $a$:
\beq
c=1+\epsilon_\psi v^2 .
\eeq
Here $v=\vev{H}$ is the Higgs vacuum expectation value. 
In sect.~\ref{secprop} we found that a space-dependent fermion mass gives a distortion to the ``light speed" already at order $a^2$. 
Instead, here we are finding that, if the Lorentz violation originates in the Higgs mass, the effect for the fermions starts only at order $a^6$. 
The reason is that, in such a case the inhomogeneity in $v$ and consequently in the fermion mass $m=\lambda v$,
is itself suppressed by two powers of $a$, due to the effect of the Higgs kinetic term.
From an effective theory point of view, the Higgs vacuum expectation value is constant and does not break Lorentz invariance. Therefore fermions can feel the effect of Lorentz violation only through higher dimensional operators, like the one in \eq{opfer}, induced by the mixing between the zero mode and the high-frequency modes of the Higgs field. 

At tree level, the Lorentz violation is communicated to gauge fields through diagrams involving Higgs and gauge particles. These diagrams have the effect of making the derivatives contained in \eq{effo} covariant under the gauge group and also generate some new higher-dimensional operators.

In summary, we have considered the effects of a space-time varying Higgs mass in the Standard Model. The source of Lorentz violation is expressed in terms of two parameters: $\mu^2$, which characterizes the amplitude of the mass square variations, and $|a|$ (or $1/\Lambda$), which defines the wavelength (or frequency) of these variations. The most appropriate language to address the problem is that of an effective low-energy field theory, valid below the scale $\Lambda$. In the effective theory, all the effects are induced by the mixing of the Higgs modes, which is proportional to $\mu^2/\Lambda^2$. The Higgs vacuum expectation value and all masses are constant in the effective theory, but the Higgs kinetic term is modified by a Lorentz-violating operator. This operator is renormalizable, but its coefficient is proportional to $\mu^4/\Lambda^4$, and thus suppressed by four powers of the cutoff scale. All other effects can be written in terms of higher-dimensional operators, suppressed by additional powers of $\Lambda$. For instance, the Lorentz-violating effects in the fermion kinetic term are proportional to $\mu^4v^2/\Lambda^6$. These conclusions are based on tree-level considerations. Now we turn to discuss the effects of quantum corrections.

\section{Lorentz violation at loop level}
Let us consider the effective Lagrangian containing the dominant rotationally invariant dimension-4 Lorentz-violating operators, in the rest frame of the Lorentz-breaking sector
\footnote{
The most generic Lorentz-breaking Lagrangian in the notations of~\cite{Kost}
can be reduced to this form setting 
$(k_{\phi\phi})_{00}=-2\delta c_H$,
$(k_F)_{i0i0}=(k_F)_{0i0i}=-(k_F)_{0ii0}=-(k_F)_{i00i}=\delta c_A$,
 $(c_\psi)_{00} =  \delta c_\psi$.
 All other tensors vanish and all other components of these tensors vanish.
In general, such tensors describe non-isotropic Lorentz violation~\cite{Kost}. 
 Performing Lorentz transformations on~\eq{eq:Lagc}
 the other components of the $c_\psi,k_F,k_{\phi\phi}$ tensors 
 are generated as dictated by their Lorentz structure.}

\beq\label{eq:Lagc}
 \Lag  = \Lag_{\rm SM} - 2\delta c_H |(\vec \nabla + i  g\vec A) H|^2 -\sum_A {\delta c_A} F_{0i}^2-
\sum_\psi \delta c_\psi ~ i \bar\psi\, \vec\gamma\cdot(\vec\nabla+i g\vec A) \psi .
\eeq
Here $A=\{Y, W^a, G^a\}$ describes the SM gauge bosons, $\psi$ are the 15 SM Weyl (chiral) fermion multiplets
($L,E,Q,U,D$, appearing in 3 generations)
 and $H$ is the scalar Higgs.
The coefficients $\delta c$ are the corrections to their `speed-of-light'.
Each of the various $\delta c$ can be set to zero by rescaling time: $t \to (1+\delta c/c) t$;
the difference between the $\delta c$ of different particles has physical meaning.
At tree-level only $\delta c_H$ is non-zero, and
it only affects the Higgs velocity and the $W,Z$ masses, which are negligibly probed by experiments.

\begin{figure}[t]
$$\includegraphics{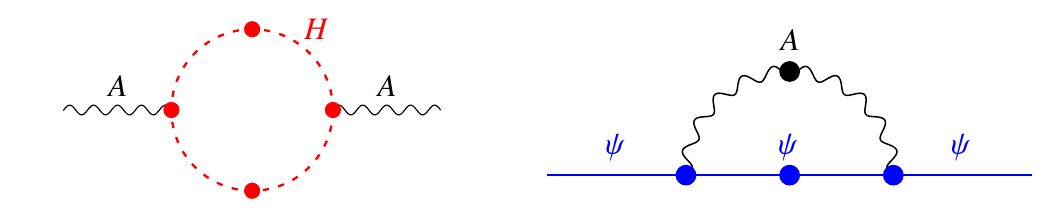}$$
\caption{\em Loop diagrams feeding the original Lorentz violation into the gauge and fermion sectors.The dots denote all possible insertions of Lorentz-violating operators, to be performed one-by-one.
\label{Feyn}}
\end{figure}

At one-loop level $\delta c_H$ induces a correction to the speed-of-light of
SU(2)$_L$ and hypercharge electroweak vectors $A$
The propagator of these vectors, taking into account the one loop correction of \fig{Feyn}a, is
\beq -i\Pi_{\mu\nu} =  [p_\mu p_\nu - p^2 \eta_{\mu\nu}]_{c_A}
- \frac{2b_Hg^2_A}{(4\pi)^2} \left( \frac{1}{\epsilon}+\ln \frac{\mu}{\Lambda}\right)   [p_\mu p_\nu - p^2 \eta_{\mu\nu}]_{c_H},
\label{eq:Pi}
\eeq
that differs from the standard expression only because we specified the speed of light to be used in 
the various terms, e.g.\ $p^2 = E^2/c^2 -  {\vec p}^2$.
The group theory coefficient is 
$b_H=1/6$ for both SU(2)$_L$ and U(1)$_Y$ (normalized such that the $H$ hypercharge is $1/2$).
Thereby
\beq \delta c_2 \simeq  \frac{g_2^2}{48\pi^2}\delta c_H \ln\frac{\Lambda}{m_h},\qquad
\delta c_1 \simeq \frac{g_Y^2}{48\pi^2}\delta c_H \ln\frac{\Lambda}{m_h}
\eeq
where $m_h$ is the physical Higgs mass
and $ c_\gamma =c_1 \cos^2\theta_{\rm W} + c_2 \sin^2\theta_{\rm W}$ for the photon.
After a further loop correction, one also gets a $\delta c$ for the SM fermions  (\fig{Feyn}b):
\beq \delta c_\psi \sim  \delta c_{1,2} \frac{g^2}{(4\pi)^2}  \ln\frac{\Lambda}{m_h} .\eeq

\subsection{RGE for the speed-of-light}
The loop effects are best described by a system of RGE for the speed-of-light of the various SM particles, that allows us as usual to re-sum the log-enhanced
corrections.

We consider a generic theory with particles $p$
(gauge vectors $A$, Weyl fermions $\psi$ and scalars $H$) interacting
among them with gauge couplings $g_A$, Yukawa couplings $\lambda$. 
The quartic scalar couplings do not enter in the RGE under consideration.
We find that the RGE equations for their maximal speeds
are:\footnote{RGE equations for Lorentz-violating tensors have been already computed for QED in ref.~\cite{LVRGE}.
Renormalizability of theories with higher-dimensional Lorentz-violating operators has been studied in~\cite{Anselmi}.
}
\begin{eqnsystem}{sys:RGE}
(4\pi)^2 \frac{d\, c_A}{d\ln\mu} &=& 2 g_A^2 \sum_p b_p (c_A-c_p )\\
(4\pi)^2 \frac{d\, c_\psi}{d\ln\mu} &=&\frac{16}{3}
\sum_A  g_A^2 C_A(c_\psi-c_A) + \sum_{\psi',H} \lambda_{\psi\psi'H}^2 
(\frac{c_\psi}{2} - \frac{c_{\psi'}}{6} - \frac{c_H}{3})\\
(4\pi)^2 \frac{d\, c_H}{d\ln\mu} &=&4
\sum_A  g_A^2 C_A(c_H-c_A) +\sum_{\psi,\psi'} \lambda_{\psi\psi'H}^2 
(2c_H- c_\psi - c_{\psi'})
\end{eqnsystem}
where $b_p$ are the well known coefficients that enter
in the RGE for the gauge couplings:
\beq(4\pi)^2 \frac{d g_A}{d\ln\mu} =   b_A g_A^3 \qquad
b_A = \sum_p b_p = -\frac{11}{3} T_1^2 + \frac{2}{3} T^2_{1/2} + \frac{1}{3} T_0^2.\eeq
The group factors are defined as  $\hbox{Tr}\, T^a T^b = T^2\delta^{ab}$ ($T^2=1/2$ for the fundamental of SU($n$),
and $T^2=n$ for the adjoint) and as $  (T^a_A T^a_A)_{ij} = C_A \delta_{ij}$ ($C_1=q^2$ for a U(1) charge;
$C_2=3/4$ for the fundamental of SU(2); $C_3=4/3$ for the fundamental of SU(3)).
The self-renormalization terms in eqs.~(\ref{sys:RGE}) come from wave-function renormalizations.
Note that the Lorentz-invariant limit $c_p=c$ is RGE invariant.

In the Standard Model, summing over the three generations of $E,L,U,D,Q$
and using the GUT normalization $g_1 = \sqrt{5/3} g_Y$, the RGE are:
\begin{eqnsystem}{sys:RGESM}
(4\pi)^2 \frac{d\, c_H}{d\ln\mu} &=&\frac{3}{5}g_1^2(c_H-c_1)+3g_2^2 (c_H-c_2)\\
(4\pi)^2 \frac{d\, c_3}{d\ln\mu} &=& g_3^2  [8c_3 - 4c_Q -2c_U -2c_D]\\
(4\pi)^2 \frac{d\, c_2}{d\ln\mu} &=& g_2^2  [25c_2-6c_L-18c_Q -c_H]/3\\
(4\pi)^2 \frac{d\, c_1}{d\ln\mu} &=& g_1^2  [41 c_1 -c_H - 4c_D - 12 c_E - 6 c_L - 2 c_Q - 16 c_U]/5\\
(4\pi)^2 \frac{d\, c_E}{d\ln\mu} &=& \frac{16}{5}g_1^2(c_E-c_1)\\
(4\pi)^2 \frac{d\, c_L}{d\ln\mu} &=& \frac{4}{5}g_1^2(c_L-c_1)+4g_2^2 (c_L-c_2)\\
(4\pi)^2 \frac{d\, c_Q}{d\ln\mu} &=& \frac{64}{9}g_3^2(c_Q-c_3)+\frac{4}{45}g_1^2(c_Q-c_1)+4g_2^2 (c_Q-c_2)\\
(4\pi)^2 \frac{d\, c_U}{d\ln\mu} &=& \frac{64}{9}g_3^2(c_U-c_3)+\frac{64}{45}g_1^2(c_U-c_1)\\
(4\pi)^2 \frac{d\, c_D}{d\ln\mu} &=& \frac{64}{9}g_3^2(c_D-c_3)+\frac{16}{45}g_1^2(c_D-c_1).
\end{eqnsystem}
We neglect here the effect of Yukawa couplings.
After the breaking of the electroweak symmetry, the SM fermions acquire Dirac masses $m$.
In the non-relativistic limit, the speed-of-light $c_L$ and $c_R$ of the left and right handed components
become the fermion speed-of-light $c = (c_L+c_R)/2$ plus a Lorentz-breaking Hamiltonian operator $\sim (c_L-c_R) \vec p\cdot \vec\sigma$.
At first order in perturbation theory around the Lorentz-symmetric state, $\vec p\cdot \vec\sigma$  changes the energy of one given state by an amount proportional to its matrix element, which vanishes  being odd in $\vec p$.

\begin{figure}[t]
\begin{center}
$$\includegraphics{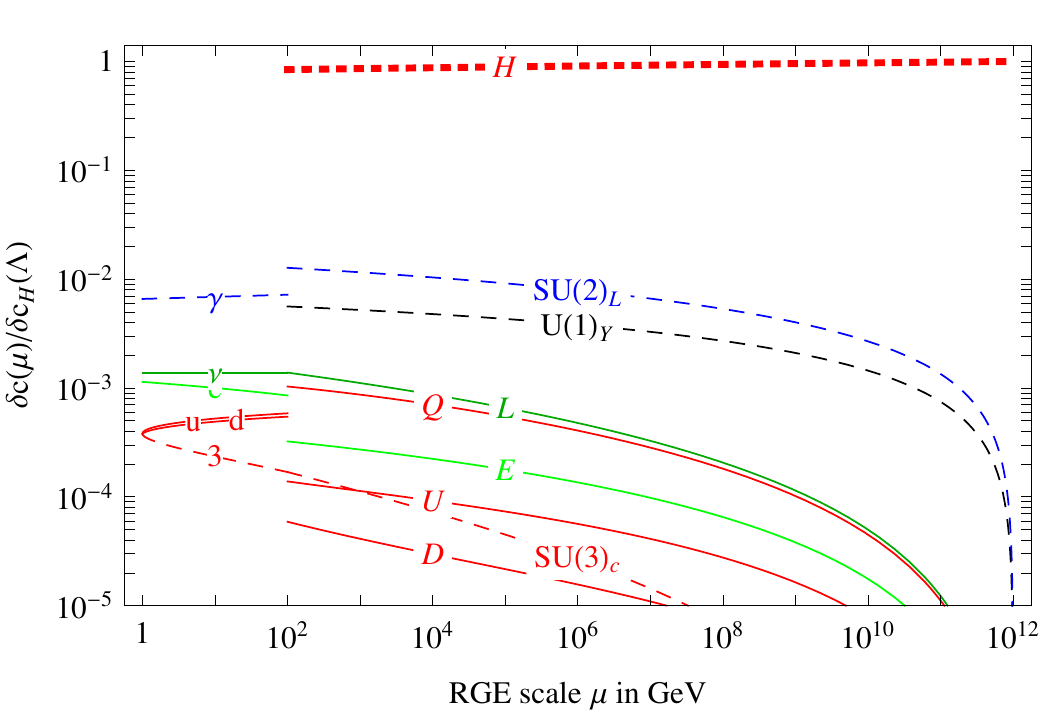}$$
\caption{\em Renormalization group evolution of the the speed-of-light of the various SM particles
from $\Lambda = 10^{12}\GeV$ to the weak scale and then down to the QCD scale.\label{fig:RGE}}
\end{center}
\end{figure}

\begin{figure}
\begin{center}
$$\includegraphics[width=0.7\textwidth]{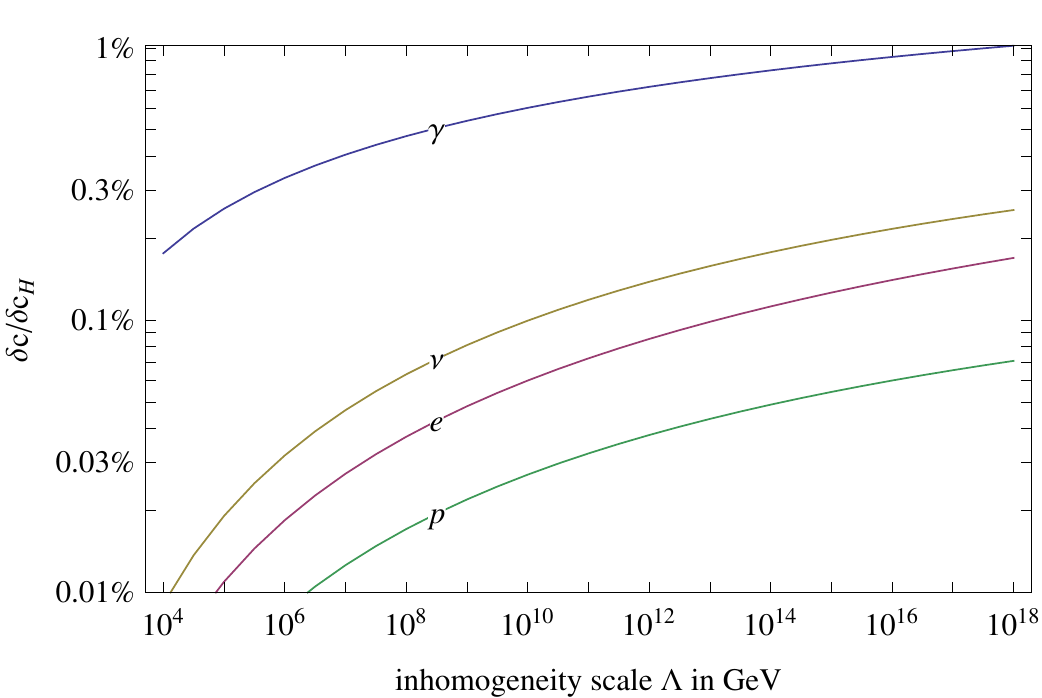}$$
\caption{\em Higgs-portal predictions for the speed-of light $\delta c_p = c_p -c$
of the stable SM particles $p=\{\gamma, e,\nu,p\}$ in units of the correction to the Higgs speed-of-light $\delta c_H \sim -\mu^4/\Lambda^4$.\label{fig:dca}}
\end{center}
\end{figure}

\section{Signals and bounds}


In sect.~3.1 we have found that Lorentz violation from the Higgs portal predicts, at tree level and at the RGE scale $\Lambda\sim 1/a$,  
a negative correction to the Higgs speed-of-light, $\delta c_H \approx - \mu^4/\Lambda^4$.
Fig.~\ref{fig:RGE} shows how the original Lorentz violation feeds into the various SM particles through the RG evolution
from $\Lambda = 10^{12}\GeV$ to the weak scale and then down to the QCD scale $\Lambda_{\rm QCD}$.

We note an interesting generic feature of the RGE: in the limit in which a gauge coupling becomes strong, all values of $c$ of particles charged under the gauge group become equal. In particular, when
the QCD coupling becomes strong, $g_3\to \infty$
at $\mu\sim \Lambda_{\rm QCD}$,
all colored particles reach a common $c$:
speed differences get exponentially suppressed by $\exp{(-k \int g_3^2 d\ln\mu})$ factors, where $k$ is a numerical constant.
To compute the common $c$, one notices that the strong coupling does not renormalize the combination
$16 c_3 + 9\sum_q c_q$ (summed over light quarks $q$).
This means that Lorentz invariance can be dynamically emergent if all SM particles felt at 
some energy a strong coupling. This could be possible in a SU(5) model such that the unified coupling runs to a large enough value.

\begin{figure}[t]
\begin{center}
$$\includegraphics[width=0.6\textwidth]{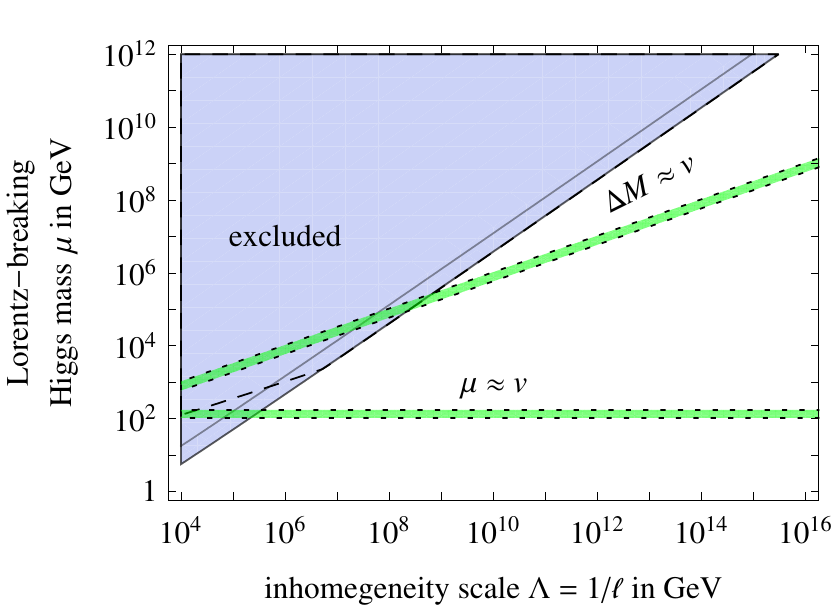}$$
\caption{\em Bounds from searches for Lorentz-violation
on the inhomogeneous Higgs mass $\mu$ and on its inhomogeneity scale $1/\ell$.
The green dotted bands indicate weak-scale-related values of the Lorentz-breaking Higgs mass $\mu$ or
of the induced effective Higgs mass.\label{fig:bounds}}
\end{center}
\end{figure}

Fig.~\ref{fig:dca} shows the predictions for the modifications of the speed-of-light of the stable SM particles in units of the correction relative to the Higgs, as functions of $\Lambda$.
The pattern is qualitatively similar for all values of $\Lambda$, and the main effect is a slower speed-of-light for the photon than for other SM particles:
$$ c_\gamma < c_\nu < c_e < c_{n,p}.$$
This pattern is mainly probed by the following observations,
and fig.~\ref{fig:bounds} summarizes the resulting bounds on the Higgs portal parameters $\Lambda=1/\ell$ and $\mu$:
\begin{itemize}
\item Proton vacuum \v{C}erenkov radiation, namely $p\to p\gamma$ decays, would 
become kinematically allowed at  $E_p > m_p /\sqrt{c_p- c_\gamma}$~\cite{CG}.
Protons have been observed in cosmic rays up to $E_p \sim 10^{7}\TeV$, and thus
$c_p - c_\gamma < 0.9\times 10^{-15}$~\cite{CG,cp}.
This bound is plotted as a dashed line:
the change in its slope arises because we only consider  cosmic ray energies below the cut-off $\Lambda$ of our theory.

\item Similarly, electron vacuum \v{C}erenkov radiation, namely $e\to e\gamma$ decays, 
would become kinematically allowed at  $E_e > m_e /\sqrt{c_e- c_\gamma}$,
but cosmic ray electrons have been observed up to 2 TeV, so that $c_e - c_\gamma< 10^{-13}$~\cite{cgamma}.

\item A stronger constraint arises from the $\gamma\gamma'\to e^-e^+$ process, in which energetic $\gamma$ are absorbed when traveling in a background of low energy $\gamma'$. The process becomes kinematically allowed if
$E'_\gamma> m_e^2/E_\gamma+ E_\gamma (c_e-c_\gamma)/2$. 
Observations of cosmic rays photons up to  $E_\gamma \sim 20\TeV$ imply
$c_e - c_\gamma < 2m_e^2/E_\gamma^2\sim 1.3\times 10^{-15}$~\cite{ce,cgamma}.

\item Furthermore, the agreement of electron synchrotron radiation at the LEP accelerator with its standard expression
implies $|c_e-c_\gamma| < 5\times 10^{-15}$~\cite{collider}.
A numerically similar bound can be deduced from astrophysical observations of
Inverse Compton and synchrotron radiation~\cite{Alts}.

\item Stability of various types of spectral lines despite the motion of the earth implies strong bounds
on Lorentz-violating operators~\cite{Heff}, 
but not on the $\delta c$ operators present in our scenario (also considered in~\cite{CG}),
at leading order in $\delta c$.
Subdominant bounds are listed in ref.~\cite{LVtab}.\footnote{Theoretical plausibility suggests
 a generic much stronger bound on Lorentz-violating scenarios, including the one we considered.
As emphasized in~\cite{Redi},
whatever breaks Lorentz invariance has an energy density which couples to gravity, but cosmological observations
suggest the presence of a Lorentz-invariant vacuum energy density, $\rho\sim $ meV$^4$.
Such a small cosmological constant poses a puzzle even to Lorentz invariant scenarios,
and the only known way out is a cancellation requiring a very fine tuning.
In presence of Lorentz breaking, such cancellations seem to need a fluid with negative energy density. 
Including quantum corrections to the vacuum energy up to experimentally probed 
energies around the weak scale $v$, one needs to impose $\delta c  ~v^4 < \rho$ 
such that $\delta c < 10^{-60}$.}

\end{itemize}
The picture also shows two more plausible values for the space-time dependent part $\mu$ of the Higgs mass:
a) $\mu$ at the weak scale;
b) $\mu$ such that its contribution to the Higgs mass in the low-energy effective theory
$\Delta M^2\sim \mu^4 \ell^2$ (negative for space-like inhomogeneities) of \eq{epsge}
is at the weak scale.  In such a case electroweak symmetry breaking could be a byproduct of inhomogeneities;
the result $\Delta M^2 \ll \mu^2$ holds because inhomogeneities in the Higgs vev are suppressed by the
Higgs kinetic term $|\partial_\mu H|^2 \sim v^2/\ell^2$.

We also considered sub-leading effects, suppressed by powers of $E/\Lambda$, which lead to variations in the speed-of-light
that depend on the energy $E$.
The main experimental constraints on such effect are:
\beq\label{c(E)}
|c_\gamma(E) - c_\gamma(E')| \lsim 
\left\{
\begin{array}{ll}
 10^{-19}\qquad & \hbox{at $E,E'\sim 0.1\GeV$~\cite{Fermi}}\cr
10^{-15}& \hbox{at $E,E'\sim\TeV$~\cite{Magic}}
 \end{array}\right.\eeq
 Such bounds are not competitive. Furthermore,
in our scenario (and actually more in general) also the dominant  effects depends on energy,
due to the logarithmic RGE running of $c$.
Since $|\delta c_H | \gg |\delta c_\gamma|$ in our model, $c_\gamma$ has a sizable RGE running
above the weak scale, $d\ln c_\gamma/d\ln \mu\sim 10^{-3}$, and a much slower running at lower
energies below Higgs decoupling, $d\ln c_\gamma/d\ln \mu\sim 10^{-5}$.
The limits in \eq{c(E)} thereby imply a bound $|\delta c_\gamma| < 10^{-12}$, which is again not competitive
with the constraints previously described.

The main qualitative point is that  Lorentz violation in the Higgs sector
must be suppressed by a scale well above the electroweak scale.
This means that various possible solutions to the Higgs mass hierarchy problem
that one can invent using  Lorentz violation (e.g.\ assuming that the Higgs is a 2d field localized
on strings that fill the space; or adding spatial gradients $|\vec\nabla H|^4$ to the Lagrangian)
are experimentally too strongly constrained to make the weak scale naturally small.

%

\section{Discussion and conclusions}
Lorentz violation is often phenomenologically studied by considering only non-renormalizable operators leading to
corrections to the speed-of-light of the form $\delta c \sim (E/\Lambda)^p$, where $p=1$ or 2
and $\Lambda$ is some high-energy scale, maybe the Planck scale.
On the theoretical side, once Lorentz symmetry is broken, one expects that the renormalizable
terms are also strongly affected (at least after that quantum corrections are taken into account), such that
there are order-unity differences in the speed-of-light of different particles, $\delta c\sim 1$, in dramatic contrast with
the experimental bound $|\delta c|< 10^{-15}$. 

In this paper we have considered a specific and well-defined source of Lorentz violation. It originates in a hidden sector and it is communicated to the Standard Model through the Higgs portal: only the Higgs mass term $\mu^2 |H|^2$ violates Lorentz invariance,
being inhomogeneous on small scales $1/\Lambda$.
This naturally leads to a small correction to the Higgs speed-of-light, $\delta c_H\sim- (\mu/\Lambda)^4$.
We computed this effect at tree level in two ways: i) in section 2 we have solved the propagation equations in a simple inhomogenous background;
ii) in section 3 we have derived an effective Lagrangian: inhomogeneities lead to mixing between low and high-frequency modes, 
so that the integration out of the high-frequency modes gives a Lorentz-violating effective operator, $|(\vec\nabla+ i g\vec{A}) H|^2$.
Fermions are affected only by higher dimension operators, which we have computed.

At loop level, $\delta c_H$ propagates to all other SM particles via a system of RGE equations for their speed-of-light;
\fig{fig:RGE} shows a typical solution.  
An interesting feature is that the strong coupling dynamically drives the speed-of-light of all colored particles to a common value.
The signals and bounds of our scheme of Lorentz violation were explored in section 4.


\paragraph{Acknowledgements} 
We thank Leonardo Giusti,
Riccardo Rattazzi, and Michele Redi for discussions.
This work was supported by the ESF 8090 and  SF0690030s09.

\small
\begin{multicols}{2}

\end{multicols}
\end{document}